\documentclass[aps,prl,preprint,tightenlines,superscriptaddress,showpacs,byrevtex]{revtex4}
\usepackage{graphicx} 
\usepackage{dcolumn}  

\graphicspath{{ps}}

\newcommand{\ee}{e^{+}e^{-}}

\newcommand{\jp}{J/\psi}

\newcommand{\psip}{\psi(2S)}

\newcommand{\pipi}{\pi^{+}\pi^{-}}
\newcommand{\piz}{\pi^{0}}
\newcommand{\bbar}{B\bar{B}}

\newcommand{\Mbc}{M_{\rm bc}}
\newcommand{\DE}{\Delta E}

\newcommand{\rt}{\rightarrow}

\newcommand{\etal}{\em et al.}

\begin{document}


\preprint{\vbox{ \hbox{   }
                   \hbox{BELLE-CONF-0541}
                   \hbox{LP2005-176} 
}}

\title{ \quad\\[0.5cm] Experimental constraints on 
the possible $J^{PC}$ quantum numbers of
the $X(3872)$ }

\affiliation{Aomori University, Aomori}
\affiliation{Budker Institute of Nuclear Physics, Novosibirsk}
\affiliation{Chiba University, Chiba}
\affiliation{Chonnam National University, Kwangju}
\affiliation{University of Cincinnati, Cincinnati, Ohio 45221}
\affiliation{University of Frankfurt, Frankfurt}
\affiliation{Gyeongsang National University, Chinju}
\affiliation{University of Hawaii, Honolulu, Hawaii 96822}
\affiliation{High Energy Accelerator Research Organization (KEK), Tsukuba}
\affiliation{Hiroshima Institute of Technology, Hiroshima}
\affiliation{Institute of High Energy Physics, Chinese Academy of Sciences, Beijing}
\affiliation{Institute of High Energy Physics, Vienna}
\affiliation{Institute for Theoretical and Experimental Physics, Moscow}
\affiliation{J. Stefan Institute, Ljubljana}
\affiliation{Kanagawa University, Yokohama}
\affiliation{Korea University, Seoul}
\affiliation{Kyoto University, Kyoto}
\affiliation{Kyungpook National University, Taegu}
\affiliation{Swiss Federal Institute of Technology of Lausanne, EPFL, Lausanne}
\affiliation{University of Ljubljana, Ljubljana}
\affiliation{University of Maribor, Maribor}
\affiliation{University of Melbourne, Victoria}
\affiliation{Nagoya University, Nagoya}
\affiliation{Nara Women's University, Nara}
\affiliation{National Central University, Chung-li}
\affiliation{National Kaohsiung Normal University, Kaohsiung}
\affiliation{National United University, Miao Li}
\affiliation{Department of Physics, National Taiwan University, Taipei}
\affiliation{H. Niewodniczanski Institute of Nuclear Physics, Krakow}
\affiliation{Nippon Dental University, Niigata}
\affiliation{Niigata University, Niigata}
\affiliation{Nova Gorica Polytechnic, Nova Gorica}
\affiliation{Osaka City University, Osaka}
\affiliation{Osaka University, Osaka}
\affiliation{Panjab University, Chandigarh}
\affiliation{Peking University, Beijing}
\affiliation{Princeton University, Princeton, New Jersey 08544}
\affiliation{RIKEN BNL Research Center, Upton, New York 11973}
\affiliation{Saga University, Saga}
\affiliation{University of Science and Technology of China, Hefei}
\affiliation{Seoul National University, Seoul}
\affiliation{Shinshu University, Nagano}
\affiliation{Sungkyunkwan University, Suwon}
\affiliation{University of Sydney, Sydney NSW}
\affiliation{Tata Institute of Fundamental Research, Bombay}
\affiliation{Toho University, Funabashi}
\affiliation{Tohoku Gakuin University, Tagajo}
\affiliation{Tohoku University, Sendai}
\affiliation{Department of Physics, University of Tokyo, Tokyo}
\affiliation{Tokyo Institute of Technology, Tokyo}
\affiliation{Tokyo Metropolitan University, Tokyo}
\affiliation{Tokyo University of Agriculture and Technology, Tokyo}
\affiliation{Toyama National College of Maritime Technology, Toyama}
\affiliation{University of Tsukuba, Tsukuba}
\affiliation{Utkal University, Bhubaneswer}
\affiliation{Virginia Polytechnic Institute and State University, Blacksburg, Virginia 24061}
\affiliation{Yonsei University, Seoul}
  \author{K.~Abe}\affiliation{High Energy Accelerator Research Organization (KEK), Tsukuba} 
  \author{K.~Abe}\affiliation{Tohoku Gakuin University, Tagajo} 
  \author{I.~Adachi}\affiliation{High Energy Accelerator Research Organization (KEK), Tsukuba} 
  \author{H.~Aihara}\affiliation{Department of Physics, University of Tokyo, Tokyo} 
  \author{K.~Aoki}\affiliation{Nagoya University, Nagoya} 
  \author{K.~Arinstein}\affiliation{Budker Institute of Nuclear Physics, Novosibirsk} 
  \author{Y.~Asano}\affiliation{University of Tsukuba, Tsukuba} 
  \author{T.~Aso}\affiliation{Toyama National College of Maritime Technology, Toyama} 
  \author{V.~Aulchenko}\affiliation{Budker Institute of Nuclear Physics, Novosibirsk} 
  \author{T.~Aushev}\affiliation{Institute for Theoretical and Experimental Physics, Moscow} 
% ...
\collaboration{The Belle Collaboration}

\noaffiliation

\begin{abstract}
We examine possible $J^{PC}$ quantum number assignments
for the $X(3872)$.  Angular correlations between
final state particles in $X(3872)\rt \pipi\jp$ decays
are used to rule out $J^{PC}$ values of 
$0^{++}$ and $0^{-+}$.
The shape of the $\pipi$ mass distribution near its
upper kinematic limit favors $S$-wave over $P$-wave
as the relative orbital angular momentum between 
the final-state dipion and $\jp$, which strongly
disfavors $1^{-+}$ and $2^{-+}$ assignments.  The accumulated
evidence strongly favors a $J^{PC}=1^{++}$ assignment
for the $X(3872)$, although the $2^{++}$ possibility
is not ruled out by tests reported here.  
The analysis is based on a sample of
$X(3872)$ mesons produced via the exclusive process 
$B\rt K X(3872)$ in a 256~fb$^{-1}$ data sample
collected at the $\Upsilon(4S)$ resonance in the Belle detector at 
the KEKB collider.

\end{abstract}


\maketitle


{\renewcommand{\thefootnote}{\fnsymbol{footnote}}}
\setcounter{footnote}{0}

The $X(3872)$ was first observed by Belle  
in exclusive  $B^-\rt K^-\pipi\jp$ 
decays~\cite{skchoi_x3872,conj}.  
The subsequent observation of
the $X(3872)\rt\gamma\jp$ decay mode~\cite{skchoi_gamjpsi}
established the charge parity as $C=+1$.
In the same paper, Belle also reported evidence for the
decay $X\rt \pipi\pi^0\jp$, where
the $\pipi\piz$ invariant mass distribution has a
strong peak between 750~MeV and the kinematic limit of 775~MeV,
suggesting that the process is dominated by the sub-threshold
decay $X\rt\omega\jp$.     The  partial widths for
$3\pi\jp$ and $2\pi\jp$ decays are of comparable size, 
which implies a large violation of isospin symmetry.

Here we report on a study of $X(3872)\rt \pipi\jp$
decays produced via the exclusive decay process $B\rt KX(3872)$. 
We use a data sample that contains 275~million $\bbar$ pairs 
collected in the Belle detector at the KEKB energy-asymmetric 
$\ee$ collider.  The data were accumulated at a center-of-mass 
system (cms) energy of $\sqrt{s} = 10.58$~GeV, corresponding to 
the mass of the $\Upsilon(4S)$ resonance.  KEKB is described 
in detail in ref.~\cite{KEKB}.

The Belle detector is a large-solid-angle magnetic 
spectrometer  that consists of a three-layer silicon vertex 
detector, a 50-layer cylindrical drift chamber (CDC), an 
array of aerogel threshold Cherenkov counters (ACC),  a 
barrel-like arrangement of time-of-flight  scintillation 
counters (TOF), and an electromagnetic calorimeter
(ECL) comprised of CsI(Tl) crystals  located inside
a superconducting solenoid coil that provides a 1.5~T
magnetic field.  An iron flux-return located outside of 
the coil is instrumented to detect $K_L$ mesons and to 
identify muons (KLM).  The detector is described in detail 
elsewhere~\cite{Belle}.

We select events that contain a $\jp$, either a charged or neutral kaon, 
and a $\pipi$ pair using criteria described in 
refs.~\cite{skchoi_x3872} and~\cite{skchoi_y3940}.  
To reduce the level of $\ee\rt q\bar{q}$ ($q=u,d,s~{\rm or}~c$-quark)
continuum events in the sample,
we also require  $R_2 < 0.4$, where $R_2$ is the normalized
Fox-Wolfram moment~\cite{fox}, and $|\cos\theta_B|<0.8$, where
$\theta_B$ is the polar angle of the $B$-meson direction
in the cms.

Candidate $B \rt K\pipi\jp$ mesons are identified by the 
energy difference  $\DE\equiv E_B^{\rm cms} - E_{\rm beam}^{\rm cms}$
and the beam-energy constrained mass 
$\Mbc\equiv\sqrt{(E_{\rm beam}^{\rm cms})^2-(p_B^{\rm cms})^2}$,
where $E_{\rm beam}^{\rm cms}$ is the cms beam
energy, and $E_B^{\rm cms}$ and $p_B^{\rm cms}$ are the cms 
energy and momentum of the $K\pipi\jp$ combination.  We select 
events with $M_{bc}>5.20$~GeV and $|\DE|<0.2$~GeV and among 
these define a signal region $5.2725~{\rm GeV} < \Mbc <5.2875$~GeV
and $|\DE |< $ 0.034 GeV; this corresponds to $\pm 3\sigma$ 
from the central values for each variable.

We select events with a dipion invariant mass requirement
of $M_{\pipi} > (M(\pipi\jp) - (m_{\jp} + 200~{\rm MeV})$,
which corresponds to $M_{\pipi}>575$~MeV for the $X(3872)$. 
This reduces misidentified $\gamma$ conversions and
combinatoric backgrounds by 36\% 
with an $X(3872)$ signal loss of 6\%.

These selection criteria isolate a very
pure sample of $696\pm26$ 
$B\rt K\psip$, $\psip\rt\pipi\jp$ events.  These
events are used as a calibration reaction to 
determine the $\Mbc$, $\DE$ and $M(\pipi\jp)$ peak 
positions and resolution values,
and for validating the Monte-Carlo (MC) acceptance
calculations.

\begin{figure}[htb]
\includegraphics[width=0.6\textwidth]{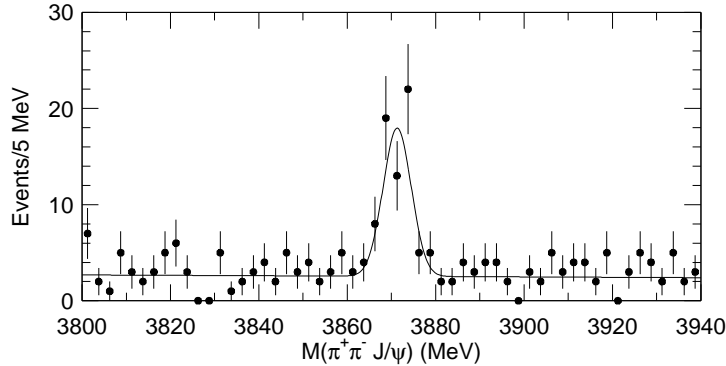}
\caption{ 
The $M(\pipi\jp)$ mass distribution 
for the $X(3872)$ region.
}
\label{fig:mpipijpsi_x3872_1box}
\end{figure}

Figure~\ref{fig:mpipijpsi_x3872_1box} shows
the $M(\pipi\jp)$ mass distribution near
3872~MeV for the selected events.   Here the 
smooth curve is the result of a fit with a 
Gaussian function to represent the $X(3872)$
signal and a first-order polynomial to 
represent the background.  The width 
of the Gaussian is fixed at $\sigma=3.2$~MeV,
the experimental resolution determined
from the $\psip\rt\pipi\jp$ event sample.
The total signal yield is
$49.1\pm 8.4$ events.  For subsequent analysis,
we define an $X(3872)$ signal region to be
$\pm 5$~MeV around the signal peak. For
background estimates, we use $\pm 50$~MeV sidebands 
above and below the signal peak
centered at 3837~MeV and 3907~MeV. 
There are a total 58 events 
in the signal region;  the background content, 
determined from the scaled sidebands, is $11.4\pm1.1$ events.

Using a MC-determined 
acceptance, we determine the
product branching fraction
$$
{\cal B}(B\rt K X(3872))\times {\cal B}(X\rt\pipi\jp)=
$$
\begin{equation}
1.31\pm 0.24 {\rm (stat)} \pm 0.13 {\rm (syst)}\times 10^{-5}.
\label{eq:pipijpsi}
\end{equation} 
where we have assumed equal $B\rt KX$ branching
fractions for charged and neutral $B$ mesons,
and that the dipion originates from $\rho\rt\pipi$.
The systematic error includes the effect 
of uncertainties in the $M(\pipi)$ shape 
for $X(3872)$ decay.  This result agrees with, and 
supersedes, the results of ref.~\cite{skchoi_x3872}.

Since both the $B$ and $K$ mesons are scalar particles,
$X(3872)$ mesons produced via exclusive $B\rt KX$ decays
cannot have a non-zero component of angular momentum along 
their  momentum direction in the $B$ rest frame.  This
provides useful limits on the number of independent
partial-wave amplitudes needed to describe the 
decay~\cite{suzuki,rosner,bugg}.

With less than fifty signal events, any
angular distribution will have, on average, only about five 
signal events per bin, which
is not sufficient for a standard angular analysis.
However, because the  signal-to-noise ratio for the 
$X\rt\pipi\jp$ signal is quite good ($S/N \simeq 4$),
a typical distribution has, on average, only about 
one or two background events per bin.  We exploit this good 
$S/N$ and try to find, for a given $J^{PC}$ hypothesis
for the $X(3872)$, angular quantities  that have 
distributions with a zero in some location.  In the bins near the 
zero point, any observed events would have to be
accounted for by upward fluctuations of the 
background~\cite{likelihood}. 

For $0^{- +}$, there is only one invariant amplitude
corresponding to a $\rho$ and $\jp$ in a $P$-wave.
The decay amplitude is proportional to the scalar triple product
of the $\rho$ and $\jp$ polarizations and their
relative momentum.  As a result, the polarizations
are perpendicular to each other and their relative momentum.
We follow a suggestion by Rosner~\cite{rosner}
and use a  coordinate system  where the $x$-axis is 
defined to be opposite the $\jp$ direction in the $\rho$
rest frame, the $x-y$ plane is defined by the $\pi^+$ and $\jp$
directions and the $z$-axis is chosen so that it forms a right-handed
coordinate system.  We define $\theta$ as the angle 
between the $\ell^+$ and the $z$ axis in the $\jp$ rest frame 
and $\psi$ as the angle between the $\pi^+$ and the $x$ 
axis in the dipion rest frame.
The expected distribution for $0^{-+}$ is
$d^2N/d(\cos\theta) d(\cos\psi) \propto \sin^2\theta\sin^2\psi$. 

\begin{figure}[htb]
\includegraphics[width=0.6\textwidth]{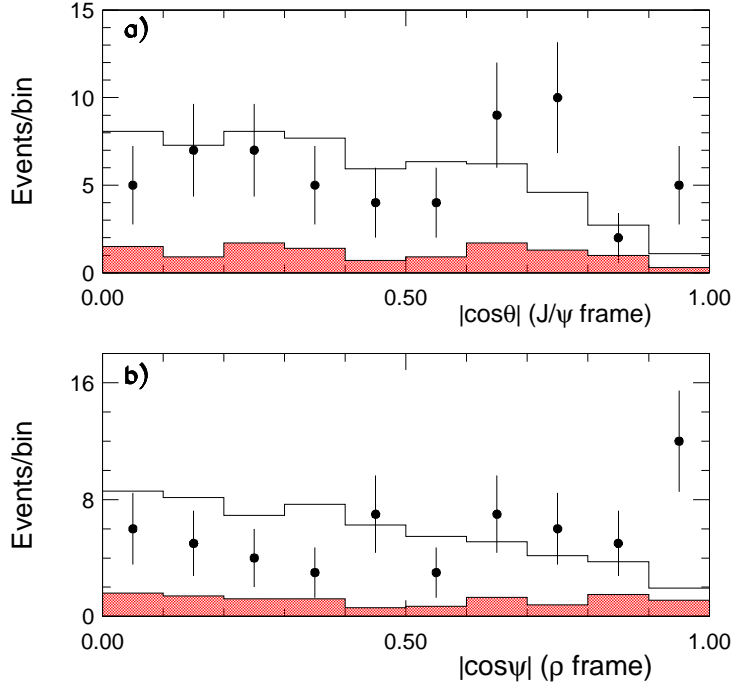}
\caption{ 
The {\bf (a)} $|\cos\theta|$
and {\bf (b)} $|\cos\psi|$ distributions for
events in the $X(3872)$ signal region (points with error bars).
The open histogram is the expected distribution for a $0^{-+}$
assignment including background.  The hatched histogram shows
the scaled sideband.
}
\label{fig:b2kx_0-+_2box_vert}
\end{figure}

The  $|\cos\theta|$ and $|\cos\psi|$ distributions
for the $X(3872)$ signal region are shown in 
Figs.~\ref{fig:b2kx_0-+_2box_vert}(a) and (b),
respectively.  The shaded histograms indicate
the side-band determined background.
The distributions for both variables
show strong signals at the upper edge
of each plot, in contrast to expectations for a
$\sin^2\theta\sin^2\psi$ dependence.  The open
histogram shows the $0^{-+}$ MC expectations 
plus background, normalized to the observed
number of events.  Here the agreement  
is marginal for $\cos\theta$: $\chi^2/d.o.f. = 17.7/9$ 
but poor for $\cos\psi$: $\chi^2/d.o.f. = 34.2/9$.
This latter distribution allows us to reject
the $0^{-+}$ assignment with high confidence.

\begin{figure}[htb]
\includegraphics[width=0.6\textwidth]{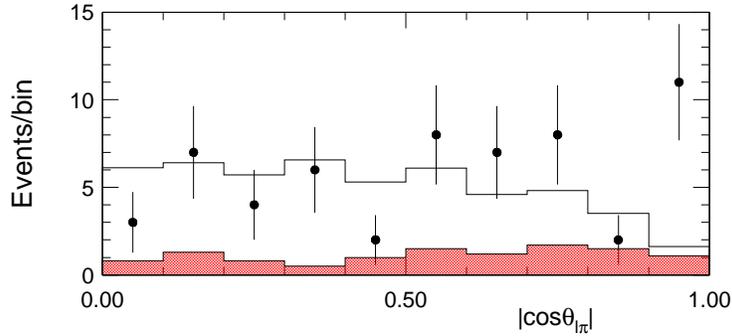}
\caption{ 
The $|\cos\theta_{\ell\pi}|$ distribution for
events in the $X(3872)$ signal region (points with error bars).
The open histogram is the expected distribution for a $0^{++}$
assignment including background.  The hatched histogram shows
the scaled sideband.
}
\label{fig:b2kx_coslpi_1box}
\end{figure}

For $0^{++}$, two invariant amplitudes are possible,
corresponding to the $\rho$ and $\jp$ in
relative $S$- or $D$-waves.  Because of the limited
phase-space, the $D$-wave 
contribution can be expected to be strongly suppressed
relative to the $S$ wave term and is ignored. The 
amplitude is then proportional to the scalar 
product of the $\rho$ and $\jp$ polarizations.
We define $\theta_{\ell\pi}$ as
the angle between the $\ell^+$ and the 
$\pi^+$ in the $X(3872)$ rest frame. 
In the limit where the $X(3872)$, $\jp$ and $\rho$ rest 
frames coincide 
$dN/d(\cos\theta_{\ell\pi})\propto\sin^2\theta_{\ell\pi} $.
The kinematic smearing due to relative motion of 
the different frames is incorporated in the MC
simulations that are used to compare data with
expectations~\cite{smearing}.

Figure~\ref{fig:b2kx_coslpi_1box} shows the
$|\cos\theta_{\ell\pi}|$ distribution, computed in
the $\rho$ rest frame, for $X(3872)$
signal region events. The agreement
with $S$-wave $0^{++}$ MC expectations 
is poor: $\chi^2/d.o.f.= 31.0/9$,  
and provides evidence against the $0^{++}$
assignment.

For $1^{++}$ the $\jp$ and $\rho$ can
be in a relative $S$ and/or $D$-wave.  We
use a coordinate system~\cite{rosner}
where the $x$-axis is the negative
of the kaon flight path, the $x-y$ plane is defined
by the kaon and $\pi^+$ and the $z$ axis completes a right-handed 
coordinate system.  
The angle between the $\pi^+$ direction
and the $x$-axis is $\chi$ and the angle  between the $\ell^+$ 
direction and the $z$-axis is $\theta_{\ell}$.  In the 
limit where the $\jp$ and $\rho$ are at rest in the $X$ rest frame
(and $D$-wave contributions can be neglected), the amplitude
is proportional to the vector triple product of 
the $X$, $\rho$ 
and $\jp$ polarizations, and the choice of axes ensures that 
the $X$ polarization is along the $x$ direction~\cite{rosner,bugg}.  
The expectation for $1^{++}$ is
$d^2N/d(\cos\theta_{\ell})d(\cos\chi)\propto\sin^2\theta_{\ell} \sin^2\chi$.

\begin{figure}[htb]
\includegraphics[width=0.6\textwidth]{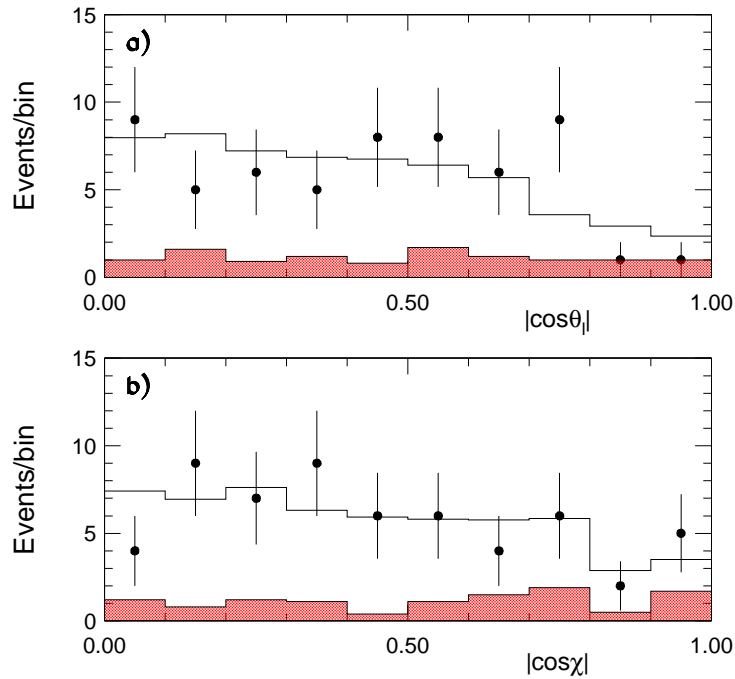}
\caption{ 
The {\bf a)}  $|\cos\theta_{\ell}|$ and {\bf b)} $|\cos\chi|$  
distribution for 
events in the $X(3872)$ signal region (points with error bars).
The open histogram is the expected distribution for a $1^{++}$
assignment including background.  The hatched histogram shows
the scaled sideband.
}
\label{fig:b2kx_1++_mc_theta_ell_coschi_2box}
\end{figure}

The $|\cos\theta_{\ell}|$ distribution 
for  $X(3872)$ signal region events is shown
in Fig.~\ref{fig:b2kx_1++_mc_theta_ell_coschi_2box}(a).
The distribution tends toward zero at the upper edge
of the plot, as expected for a $\sin^2\theta_{\ell}$ dependence.
The open histogram shows the results of
a comparison to normalized MC expectations for $1^{++}$ decaying
to a $\rho$ and $\jp$ in an $S$-wave. The agreement is 
good: $\chi^2/d.o.f=11.4/9$.
The  $|\cos\chi|$  distribution is shown in 
Figs.~\ref{fig:b2kx_1++_mc_theta_ell_coschi_2box}(b)
together with the MC expectation for $1^{++}$.
The agreement here is also good:
$\chi^2/d.o.f. = 5.0/9$.

For even-parity $C=+1$ states 
the $\pipi\jp$ final state would be a  $\rho$ and $\jp$ primarily 
in a relative $S$-wave, with some possible $D$-wave component. 
For  odd-parity states  
the $\rho$ and $\jp$ would be in a relative $P$-wave
with some possible $F$-wave.  The 
$M(\pipi)$ mass distribution near the upper kinematic 
boundary is suppressed by a $(q_{\jp}^{*})^{2\ell+1}$ 
centrifugal barrier,  where $q_{\jp}^*$ is the $\jp$ momentum
in the $X(3872)$ rest frame, and $\ell$ is the orbital
angular momentum.  For the $S$-wave ({\it i.e.} 
$J^P=J^+$) cases, the upper-boundary is modulated by
the available phase-space, which is proportional
to $q_{\jp}^*$; for a $P$-wave the modulation 
is $(q_{\jp}^{*})^3$.  Thus, the shape of the high-mass
part of the $\pipi$ invariant mass distribution
provides some $J^{PC}$ information.

\begin{figure}[htb]
\includegraphics[width=0.6\textwidth]{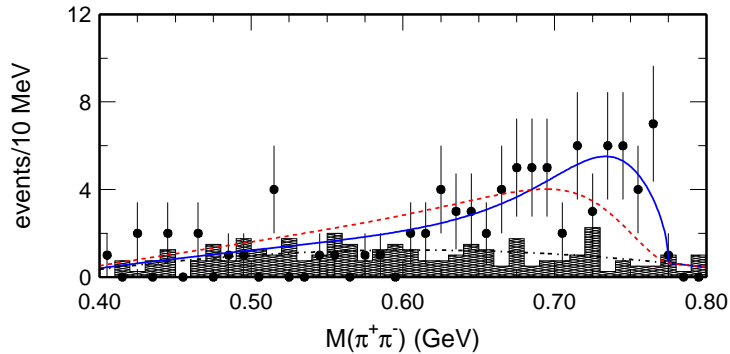}
\caption{ 
$M(\pipi)$ distribution for 
events in the $X(3872)$ signal region;
the histogram indicates the side-band determined background. 
The solid (dashed) curve shows the fit
that uses a $\rho$ Breit-Wigner line shape with
the $\jp$ and $\rho$ in a relative $S$-wave ($P$-wave).
The dot-dashed curve is a smooth parameterization of the
background that is used in the fit.
}
\label{fig:mpipi_swave_pwave_1box}
\end{figure}

Figure~\ref{fig:mpipi_swave_pwave_1box} shows the 
distribution for events in the $X(3872)\rt\pipi\jp$
signal region with the $M(\pipi)$ requirement relaxed;
the histogram indicates the side-band determined background,
which is parameterized by the fourth-order polynomial shown
in the figure as a dot-dashed curve. 
The solid curve in Fig.~\ref{fig:mpipi_swave_pwave_1box} shows 
the result of a fit to the $M(\pipi)$ distribution 
that uses the background function plus an acceptance-weighted 
$\rho$ BW line-shape with an $S$-wave cut-off factor at the upper 
kinematic boundary~\cite{pko};
the dashed curve shows the fit with a $P$-wave cut-off factor.  The
$S$-wave case fits the data well: $\chi^2/d.o.f. = 43.1/39$ (CL=28\%).
The $P$-wave fit is much poorer, $\chi^2/d.o.f. = 71.0/39$ (CL=0.1\%),
indicating that $J^{++}$ is strongly favored over $J^{-+}$.

In summary, we find that with reasonable assumptions
and a sample of $47$ $X\rt\pipi\jp$ signal events,  
we can rule out the $J^{PC} = 0^{-+}$ and $0^{++}$ assignments 
for the $X(3872)$ based on angular correlations among the
final state particles.  In addition, the $M(\pipi)$ distribution
is inconsistent with all $J^{-+}$ assignments.  

The results reported here, taken together with the 
observation of the $X(3872)\rt\gamma\jp$ decay mode~\cite{skchoi_gamjpsi},
rule out all $J^{PC}$ assignments with $J\le 2$ other 
than $1^{++}$ and $2^{++}$.
The decay angular distributions and $\pipi$ invariant
mass distribution agree well with expectations for the 
the $1^{++}$ assignment. The $2^{++}$ assignment
is not seriously challenged by any of the tests reported here,
but is made rather unlikely by Belle's recently
reported evidence for the
decay $X(3872)\rt D^0\bar{D}^{0}\piz$~\cite{gokhroo}.
The formation of $2^{++}$ from three pseudoscalars
requires at least one combination to be in a $D$-wave. Thus,
the near-threshold production of $D^0\bar{D}^{0}\piz$ 
would be suppressed by an $\ell=2$ centrifugal barrier.

The $1^{++}$ charmonium $\chi_{c1}'$ state is an unlikely
assignment for the $X(3872)$.
Potential model predictions for the $\chi_{c1}'$
mass range from 3953~MeV $\sim$ 3990~MeV~\cite{godfrey},
well above the $X(3872)$ mass.  The potential model masses 
are expected to be modified by coupling to open-charm states.  
A coupled-channel 
calculation of open-charm-induced splittings for the $\chi_{c1}'$ 
yields an upward mass shift of $+28$~MeV~\cite{quigg}.   

The decay $\chi_{c1}' \rt\pipi\jp$ would proceed via $\rho\jp$ 
and violate isospin.  The only well established isospin-violating
hadronic transition in the charmonium system is
$\psip\rt\pi^0\jp$, which has a measured partial width of
$\Gamma(\psip\rt\piz\jp)=0.27\pm 0.06$~keV~\cite{PDG}. 
This is small compared to the expected 
total width of an $M=3872$~MeV $\chi_{c1}'$ of
more than 1~MeV~\cite{godfrey,quigg}. 
A decay mode with a partial width this small 
would thus have a branching fraction that is less 
than 0.1\%.  This contradicts the recent BaBar
90\% confidence lower limit of 
${\cal B}(X(3872)\rt\pipi\jp) > 4.3\%$~\cite{coleman}.
Godfrey and Barnes calculate a partial width for
an $M=3872$~MeV $\chi_{c1}'$ to be 11~KeV~\cite{godfrey}, 
more than
an order-of-magnitude larger than that for the isospin 
violating $\psip\rt\piz\jp$ transition.  Thus, one 
expects the $\gamma\jp$ decay to be stronger 
than $\rho\jp$.  This is contradicted by our measurement: 
$\Gamma(X(3872)\rt\gamma\jp)/
\Gamma(X(3872)\rt\pipi\jp)=0.14\pm0.05$~\cite{skchoi_gamjpsi}.

The $1^{++}$ assignment is favored by models 
that treat the $X(3872)$ as a molecule-like $D^0\bar{D}^{*0}$
bound state~\cite{tornqvist,swanson_1}.  These models predict
strong isospin violations and a $\gamma\jp$ branching fraction
that is much less than that for 
$\pipi\jp$~\cite{swanson_2}, in agreement with observations.

We thank the KEKB group for the excellent operation of the
accelerator, the KEK Cryogenics group for the efficient
operation of the solenoid, and the KEK computer group and
the National Institute of Informatics for valuable computing
and Super-SINET network support. We acknowledge support from
the Ministry of Education, Culture, Sports, Science, and
Technology of Japan and the Japan Society for the Promotion
of Science; the Australian Research Council and the
Australian Department of Education, Science and Training;
the National Science Foundation of China under contract
No.~10175071; the Department of Science and Technology of
India; the BK21 program of the Ministry of Education of
Korea and the CHEP SRC program of the Korea Science and
Engineering Foundation; the Polish State Committee for
Scientific Research under contract No.~2P03B 01324; the
Ministry of Science and Technology of the Russian
Federation; the Ministry of Education, Science and Sport of
the Republic of Slovenia; the National Science Council and
the Ministry of Education of Taiwan; and the U.S.\
Department of Energy.


\begin{thebibliography}{99}
\bibitem{conj} The inclusion of charge-conjugate modes
is always implied.
\bibitem{likelihood} We follow the PDG-recommended 
procedure for hypothesis testing with
small statistics samples and use $\chi^2= - 2\ln\lambda$,
where $\lambda$ is
defined in Eq.~32.12 of the PDG review~\cite{PDG}.
The distribution of this statistic is found to approach
the asymptotic limit faster than other $\chi^2$
statistics.
\bibitem{smearing}
The $\cos\theta_{\ell\pi}$ distributions
calculated in the $X$, $\jp$ and $\rho$ rest frames have
similar shapes;  the difference distributions 
have an rms spead of $\simeq 0.2$.
\end{thebibliography}
\end{document}